\documentclass[journal]{IEEEtran}

\usepackage[dvipsnames]{xcolor}
\usepackage[acronym]{glossaries-extra}
\usepackage[colorlinks, citecolor=blue, linkcolor=blue]{hyperref}
\usepackage[textwidth=1.6cm]{todonotes}
\usepackage{comment}
\usepackage{nameref}
\usepackage{subcaption}
\usepackage{multicol}
\usepackage{multirow}
\usepackage{rotating}
\usepackage{stfloats}
\usepackage[noadjust]{cite}
\usepackage{float}
\usepackage{amsfonts}
\usepackage{upgreek}
\usepackage{siunitx}
\usepackage{mathtools}
\usepackage{scalerel}
\usepackage{braket}
\usepackage{soul}
\usepackage{tcolorbox}
\usepackage{lipsum}

\DeclareSIUnit{\sym}{sym}
\DeclareSIUnit{\baud}{Baud}

\newcommand{\Z}[1]{\hspace{-#1pt}}

\newcommand{\Sint}[1]{S^{\mathrm{Int}}_{#1}}
\newcommand{\Pint}[1]{P^{\mathrm{Int}}_{#1}}
\newcommand{\Pnli}[1]{P^{\mathrm{Nli}}_{#1}}
\newcommand{\Real}[1]{\mathbb{R}\left\{ #1\right\}}
\newcommand{\fch}{\Delta f_{\mathrm{ch}}}
\newcommand{\Lpk}{L_{\mathrm{Pk}}}

\newcommand{\eff}{\mathrm{eff}}

\newcommand{\Nsig}{N^{\mathrm{TX}}}
\newcommand{\Nrec}{N^{\mathrm{RX}}}

\newcommand{\bNint}{\bar{N}^{\mathrm{Int}}}

\newcommand{\bNdc}{\bar{N}^{\mathrm{DC}}}
\newcommand{\Bch}{B_{\mathrm{ch}}}
\newcommand{\Bdet}{B_{\mathrm{det}}}
\newcommand{\Tdet}{\tau_{\mathrm{det}}}
\newcommand{\Rdet}{R_{\mathrm{det}}}
\newcommand{\Bel}{B_{\mathrm{el}}}
\newcommand{\etaDV}{\eta_{\mathrm{RX}}}
\newcommand{\Rv}{R_{\mathrm{V}}}
\newcommand{\Plo}{P_{\mathrm{LO}}}
\newcommand{\vel}{v_{\mathrm{el}}}
\newcommand{\rdv}{r_{\mathrm{BB84}}}
\newcommand{\rcv}{r_{\mathrm{GMCS}}}
\newcommand{\betaEC}{\beta_{\mathrm{EC}}}

\newcommand{\diag}{\mathrm{dg}}
\newcommand{\eig}{\mathrm{eig}}
\newcommand{\der}{\mathrm{d}}

\newcommand{\appref}[1]{\hyperref[#1]{Appendix~\ref*{#1}}}

\makeatletter
\newcommand{\subalign}[1]{%
  \vcenter{%
    \Let@ \restore@math@cr \default@tag
    \baselineskip\fontdimen10 \scriptfont\tw@
    \advance\baselineskip\fontdimen12 \scriptfont\tw@
    \lineskip\thr@@\fontdimen8 \scriptfont\thr@@
    \lineskiplimit\lineskip
    \ialign{\hfil$\m@th\scriptstyle##$&$\m@th\scriptstyle{}##$\hfil\crcr
      #1\crcr
    }%
  }%
}
\makeatother

\setabbreviationstyle[acronym]{long-short}

\newacronym{sdm}{SDM}{space-division multiplexing}
\newacronym{psd}{PSD}{power spectral density}
\newacronym{wdm}{WDM}{wavelength-division multiplexed}
\newacronym{skr}{SKR}{secret key rate}
\newacronym{smf}{SMF}{single-mode fiber}
\newacronym{qkd}{QKD}{quantum-key distribution}
\newacronym{ase}{ASE}{amplified spontaneous emission}
\newacronym{fwm}{FWM}{four-wave-mixing}
\newacronym{sprs}{SpRS}{spontaneous Raman scattering}
\newacronym{qber}{QBER}{quantum bit error rate}
\newacronym{gmcs}{GMCS}{Gaussian-modulated coherent-states}
\newacronym{dv-qkd}{DV-QKD}{discrete-variable QKD}
\newacronym{cv-qkd}{CV-QKD}{continuous-variable QKD}
\newacronym{air}{AIR}{achievable information rate}

\begin{document}
\title{Secret Key Rate Limits in Coexisting Classical–Quantum Optical Links}
\author{Lucas~Alves~Zischler,~\IEEEmembership{Student~Member,~IEEE,}%
        ~Amirhossein~Ghazisaeidi,~\IEEEmembership{Senior~Member,~IEEE,~Fellow,~Optica,}
        ~Antonio~Mecozzi,~\IEEEmembership{Fellow,~IEEE,~Optica,}
        and~Cristian~Antonelli,~\IEEEmembership{Senior~Member,~IEEE,~Fellow,~Optica}
\thanks{Manuscript received XXX xx, XXXX; revised XXXXX xx, XXXX; accepted XXXX XX, XXXX. Funded by the European Union (Grant Agreement 101120422). Views and opinions expressed are, however, those of the author(s) only and do not necessarily reflect those of the European Union or REA. Neither the European Union nor the granting authority can be held responsible for them. (Corresponding Author: Lucas~Alves~Zischler)}%
\thanks{Lucas~Alves~Zischler, Antonio~Mecozzi, and Cristian~Antonelli are with the Department of Physical and Chemical Sciences, University of L’Aquila, 67100 L’Aquila, Italy: (\mbox{e-mail: lucas.zischler@univaq.it}).}%
\thanks{Amirhossein~Ghazisaeidi is with the Nokia Bell Labs, 91300 Massy, France.}}%

\markboth{}
{Secret Key Rate Optimization in Coexisting Classical–Quantum Optical Links}

\maketitle

\begin{abstract}
  Classical-quantum coexistence enables cost-effective transmission of data and quantum signals over the same fiber-optic channel. Nevertheless, weak \gls*{qkd} signals are susceptible to non-linear interference generated from the classical traffic, primarily \gls*{sprs} and \gls*{fwm}, as well as to unfiltered noise. In \glsxtrshort*{qkd} protocols, increased channel loss and excess noise both reduce the \glspl*{skr}, as illustrated in this work for the two-state BB84 and \gls*{gmcs} protocols. In this study, we derive closed-form expressions for evaluating the accumulated interference power from coexisting classical signals in a quantum frequency channel. Our model enables effective design of classical-quantum systems in \glspl*{smf}, capturing the evolution of interference arising from the relevant physical phenomena. We utilize the model to examine frequency allocation in multiband transmission systems, demonstrating that, contrary to common practice of allocating \gls*{qkd} channels in the O-band, increased \gls*{skr} is achieved by placing quantum channels in the upper E-/lower S-band across the relevant scenarios.
\end{abstract}

\glsresetall


\glsresetall

\section{Introduction}

\IEEEPARstart{Q}{uantum-key} distribution (\glsxtrshort*{qkd}) enables secure key exchange, guaranteed by information limits at the quantum level~\cite{lo2014secure}. With the rising prospect of quantum computers capable of compromising widely deployed asymmetric cryptographic schemes, \glsxtrshort*{qkd} has attracted growing interest from both academia and industry as a means to safeguard deployed networks~\cite{peev2009secoqc,sasaki2011field,wang2014field,bacco2019field}.

Among current research efforts, the majority of experimental deployments transmit quantum signals over optical fiber links. Given the similar nature between classical optical communications and \glsxtrshort*{qkd}, it is logical to consider the coexistence of both applications within the same fiber network. Such coexistence eliminates the need for dedicated fibers and reduces deployment costs, and its feasibility has been extensively explored~\cite{townsend1997simultaneous,peters2009dense,eraerds2010quantum,huang2015continuous,eriksson2018joint,mao2018integrating,beppu2025coexistence}. Nevertheless, the classical channels can severely impair the weak quantum signals.

In \gls*{wdm} coexistence schemes, the dominant impairments arise from linear effects, such as unfiltered \gls*{ase}, and from non-linear phenomena, primarily \gls*{sprs} and \gls*{fwm}~\cite{kawahara2011effect,peters2009dense}. While linear noise can, in principle, be mitigated through sufficient filtering, non-linear effects are unavoidable in silica-core fibers. Although emerging hollow-core fiber technologies offer opportunities for coexistence with negligible non-linear interference~\cite{alia2022dv}, conventional \glspl*{smf} currently remain the most cost-effective and widely available solution.

The performance of a quantum channel is quantified by its \gls*{skr}, which depends not only on the quantum transmitter and receiver parameters, but also on fiber attenuation and the noise generated from the classical channels. Several prior works have investigated the impact of classical channel parameters on quantum performance. In particular, \cite{peters2009dense} presents analytical models for \gls*{sprs} and \gls*{fwm} interference, showing strong agreement with a coexistence experiment involving two classical channels and a BB84-encoded quantum signal, and demonstrating that \gls*{fwm} becomes negligible with sufficient channel spacing. In~\cite{kawahara2011effect}, the authors analyze \gls*{sprs}-induced impairments and validate their model through experimental measurements of the \gls*{sprs} efficiency. In~\cite{du2020impact}, the authors explicitly focuses on \gls*{fwm} impairments, combining analytical modeling with experimental measurements.

In this work, we provide closed-form analytical expressions to estimate the interference noise arising from all major sources induced by classical transmission, providing a comprehensive framework to aid practical implementation. The model accounts for unfiltered linear leakage at the classical transmitter, including both co-propagating contributions and counter-propagating noise generated by Rayleigh backscattering, as well as non-linear impairments due to \gls*{sprs} and \gls*{fwm}. This study builds on our previous work~\cite{zischler2025accurate}, where we derived a comprehensive semi-analytical model for estimating coexistence-induced interference that has been experimentally validated in~\cite{zischler2026experimental}. The developed model is employed to analytically evaluate the behavior of each impairment effects across different operating conditions. It is subsequently applied to quantify the impact of the fiber parameters and coexistence-induced noise on the asymptotic \gls*{skr} of representative \glsxtrshort*{qkd} protocols, namely the two-state BB84 and \gls*{gmcs} schemes, as examples of \gls*{dv-qkd} and \gls*{cv-qkd}, respectively. With the derived model, we investigate how the \gls*{skr} can be optimized when the quantum channel wavelength is a free design parameter. In multi-band coexistence transmission, \gls*{qkd} channels are commonly assigned to the O-band, while classical traffic is allocated in the C-band~\cite{townsend1997simultaneous,mao2018integrating,thomas2024quantum}, since this spectral separation provides sufficient spacing to mitigate non-linear impairments and equipment is readily available for these regions. Nevertheless, we find that \glspl*{skr} can be improved by allocating the quantum channel to the upper E-band or lower S-band, where the attenuation penalty is reduced. As \glspl*{skr} are much smaller than the transmitted classical traffic, the transmission of multiple concurrent \gls*{qkd} signals may be desirable in order to avoid key exhaustion. The similarity of \gls*{cv-qkd} systems to classical transmission devices naturally enables \gls*{wdm} of quantum signals, as demonstrated in~\cite{brunner2023demonstration}. Therefore, we also investigate the optimal continuous window for \gls*{qkd}, reaching similar agreement to the single-channel scenario. Furthermore, in multi-band classical transmission, assigning the additional classical channels to the L-band yields reduced noise in the E- and S-bands, without compromising classical transmission capacity.

\section{Analytical model and evaluation of coexistence impairments}
\label{sec:analytical}

The quantum signal exhibits negligible non-linear interaction with the classical channels due to its low optical power. Consequently, only non-linear phenomena generated exclusively by classical signals are relevant. In addition, the transmission of quantum signals does not allow optical amplification.\footnote{The discussion in this work focuses on key exchange between two peers in a single-span scenario. Some approaches extend the reach of \glsxtrshort*{qkd} by employing trusted nodes to stablish intermediate keys~\cite{sasaki2011field}, and more sophisticated entanglement-based protocols consider an intermediate node in the quantum-key exchange~\cite{scherer2011long}.} As a result, only single-span systems are considered. Furthermore, since the quantum transmitter may be located at either end of the fiber, counter-propagating scattering effects must also be accounted.

In our previous work, we introduced a comprehensive semi-analytical model encompassing the most relevant coexistence impairments, including spatial crosstalk in novel fiber types for \gls*{sdm} transmission. The general expressions for the co-propagating and counter-propagating interference are given in~\cite[Eq.~(10)]{zischler2025accurate} and~\cite[Eq.~(33)]{zischler2025accurate}, respectively. By restricting the analysis to the single-mode fiber scenario and neglecting double-backscattering, the non-linear interference noise experienced by a channel allocated at frequency $f_{i}$ evolves as
\begin{equation}
  \pm\Z{1}\frac{\der \Pnli{i}\Z{6}}{\der z}\Z{3}=\Z{4}\begin{cases}
    \Z{3}\overbrace{-\alpha_{i}\Pnli{i}}^{\text{Loss}}\Z{3}+\Z{4}\overbrace{\sum\limits_{h}\Z{2}\eta_{ih}P_{h}e^{\Z{3}-\alpha_{h}\Z{1}z}}^{\text{SpRS}}\Z{4}+\Z{1}\overbrace{\bar{\gamma}_{i}^{2}\Z{10}\sum\limits_{\substack{h,k,l\neq i\\k=h+l-i}}\Z{12}P_{h}P_{k}P_{l}\rho_{ihkl}}^{\text{FWM}},\Z{6}&\text{Fwd},\\
    \Z{3}-\alpha_{i}\Pnli{i}\Z{3}+\Z{4}\sum\limits_{h}\Z{2}\eta_{ih}P_{h}e^{\Z{3}-\alpha_{h}\Z{1}z},&\text{Bwd},\\
  \end{cases}
  \label{eq:peqs}
\end{equation}
where the left-hand sign and the labels ``Fwd'' and ``Bwd'' refer, respectively, to the co- and counter-propagating noise with respect to the classical signal direction. The $z$-dependence of the variables is omitted for conciseness. For the frequency $f_i$, the coefficient $\alpha_i$ denotes the frequency-dependent attenuation, and ${\bar{\gamma}_i=\tfrac{8}{9}\gamma_i}$ is the effective non-linear coefficient, where $\gamma_i$ is the single-polarization non-linear efficiency and the factor $\tfrac{8}{9}$ is the Manakov factor~\cite{wai1996polarization}. The coefficient $\eta_{ih}$ is the \gls*{sprs} efficiency from $f_h$ to $f_i$, $P_h$ denotes the classical signal launch power at $f_h$, and $\rho_{ihkl}(z)$ is the \gls*{fwm} efficiency given by
\begin{equation}
  \rho_{ihkl}\Z{3}=\Z{3}\begin{cases}
    \Z{1}\zeta_{h}\Real{\frac{1-e^{-\left(\frac{1}{2}\Delta\alpha_{ihkh}+j\Delta\beta_{ihkh}\right)z}}{\Delta\alpha_{ihkh}+j2\Delta\beta_{ihkh}}}e^{(\Delta\alpha_{ihkh}-\alpha_{i})z},\Z{8}&h\Z{2}=\Z{2}l,\\
    \Z{1}4\Real{\frac{1-e^{-\left(\frac{1}{2}\Delta\alpha_{ihkl}+j\Delta\beta_{ihkl}\right)z}}{\Delta\alpha_{ihkl}+j2\Delta\beta_{ihkl}}}e^{(\Delta\alpha_{ihkl}-\alpha_{i})z},\Z{8}&h\Z{2}\neq\Z{2} l,
  \end{cases}
\end{equation}
where $\Real{\cdot}$ is the real part, $\zeta_h$ is the kurtosis of the signal allocated at $f_h$~\cite[Table~I]{semrau2019modulation}, and $\Delta\alpha_{ihkl}$ and $\Delta\beta_{ihkl}$ are given by
\begin{equation}
  \begin{split}
    \Delta\alpha_{ihkl}&=\alpha_{i}-\alpha_{h}-\alpha_{k}-\alpha_{l},\\
    \Delta\beta_{ihkl}&=\beta_{i}-\beta_{h}+\beta_{k}-\beta_{l}.
  \end{split}
  \label{eq:dalphadbeta}
\end{equation}

The coefficient $\Delta\beta^{(n)}_{ihkl}$ can be approximated by its Taylor series
\begin{equation}
  \Delta\beta_{ihkl}\approx 2\pi^{2}\beta_{2,i}\left(f^{2}_{i}-f^{2}_{h}+f^{2}_{k}-f^{2}_{l}\right),
  \label{eq:dbetafull}
\end{equation}
where $\beta_{2,i}$ is the frequency-dependent group-velocity-dispersion parameter. For evenly spaced channels, Eq.~\eqref{eq:dbetafull} can be further simplified to
\begin{equation}
  \Delta\beta_{ihkl}\approx 2\pi^{2}\beta_{2,i}\left(i^{2}-h^{2}+k^{2}-l^{2}\right)\fch^{2},
  \label{eq:dbetaidx}
\end{equation}
where $\fch$ is the channel spacing.

In addition to non-linear interference, without sufficient filtering, a non-negligible amount of linear leakage with optical power $P_i$ at the quantum frequency $f_i$ can be expected. In a counter-propagation scenario, the leaked interference can backpropagate toward the quantum receiver via Rayleigh backscattering.\footnote{Backscattered \gls*{fwm} via Rayleigh scattering is possible. However, \gls*{fwm} is typically only relevant within the classical transmission band, where backscattered \gls*{sprs} is significant, as shown further in this work. By contrast, linear leakage from the classical transmitter can occur at any frequency, therefore, it is included here.}

Noting that the interference powers are sufficiently weak, such that their fields do not interact non-linearly, closed-form solutions to Eq.~\eqref{eq:peqs} can be obtained for both propagation directions. Accounting also for a small linear leakage from the classical transmitter into the quantum channel, a closed-form expression for the total noise variance at $f_i$ produced by the classical transmission is given by
\begin{equation}
  \Pint{i}\Z{3}=\Z{3}\begin{cases}
    \sum\limits_{h\neq i}\Z{1}\eta_{ih}P_{h}\Omega_{ih}\Z{2}+\Z{1}\bar{\gamma}^{2}_{i}\Z{12}\sum\limits_{\substack{h,k,l\neq i\\k=h+l-i}}\Z{12}P_{h}P_{k}P_{l}\chi_{ihkl}\Z{2}+\Z{2}P_{i}e^{-\alpha_{i}L},&\text{Fwd},\\
    \sum\limits_{h\neq i}\Z{1}\eta_{ih}P_{h}\Omega_{ih}\Z{2}+\Z{2}\Gamma_{i}P_{i}\frac{1-e^{-2\alpha_{i}L}}{2\alpha_{i}},&\text{Bwd},
  \end{cases}
  \label{eq:cf}
\end{equation}
where $L$ is the link length, $\Gamma_i$ is the frequency-dependent Rayleigh backscattering efficiency, $\Omega_{ih}$ accounts for the longitudinal evolution of the \gls*{sprs} contribution
\begin{equation}
  \Omega_{ih}=\begin{cases}
    \frac{e^{(\alpha_{i}-\alpha_{h})L}-1}{\alpha_{i}-\alpha_{h}}e^{-\alpha_{i}L},&\text{Fwd},\\
    \frac{1-e^{-(\alpha_{i}+\alpha_{h})L}}{\alpha_{i}+\alpha_{h}},&\text{Bwd},
  \end{cases}
\end{equation}
and $\chi_{ihkl}$ is the corresponding term for \gls*{fwm}, given by
\begin{equation}
  \chi_{ihkl}\Z{3}=\Z{3}\begin{cases}
    \zeta_{h}\frac{e^{\Delta\alpha_{ihkh}L}-2e^{\frac{1}{2}\Delta\alpha_{ihkh}L}\cos(\Delta\beta_{ihkh}L)+1}{\Delta\alpha^{2}_{ihkh}+4\Delta\beta^{2}_{ihkh}}e^{-\alpha_{i}L},\Z{6}&h=l,\\
    4\frac{e^{\Delta\alpha_{ihkl}L}-2e^{\frac{1}{2}\Delta\alpha_{ihkl}L}\cos(\Delta\beta_{ihkl}L)+1}{\Delta\alpha^{2}_{ihkl}+4\Delta\beta^{2}_{ihkl}}e^{-\alpha_{i}L},\Z{6}&h\neq l.
  \end{cases}
  \label{eq:fwmactual}
\end{equation}

As shown in~\cite[Fig.~7]{niu2018optimized} and~\cite[Fig.~3]{zischler2025accurate}, the \gls*{fwm} component exhibits rapid longitudinal oscillations. As the oscillation period is proportional to $\Delta\beta_{ihkl}$, the interaction among many frequency terms produces an averaging effect. By linearly averaging the \gls*{fwm} coefficients, a simplified approximation can be obtained, given by
\begin{equation}
  \chi_{ihkl}\approx\tilde{\chi}_{ihkl}\Z{3}=\Z{3}\begin{cases}
    \zeta_{h}\frac{e^{\Delta\alpha_{ihkh}L}+1}{\Delta\alpha^{2}_{ihkh}+4\Delta\beta^{2}_{ihkh}}e^{-\alpha_{i}L},&h=l,\\
    4\frac{e^{\Delta\alpha_{ihkl}L}+1}{\Delta\alpha^{2}_{ihkl}+4\Delta\beta^{2}_{ihkl}}e^{-\alpha_{i}L},&h\neq l.
  \end{cases}
  \label{eq:fwmavg}
\end{equation}

\begin{figure}[!t]
    \centering
    \includegraphics{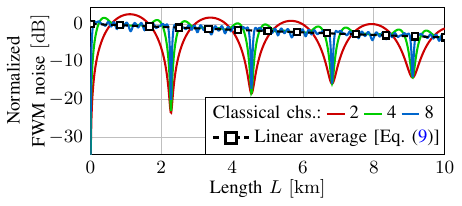}
    \caption{Normalized \gls*{fwm} noise versus fiber length for the given numbers of classical channels (color-coded), together with the linearly averaged value, shown with square markers.}
    \label{fig:FwmAverage}
\end{figure}

Figure~\ref{fig:FwmAverage} shows the relative \gls*{fwm} longitudinal intensity for an increasing count of classical channels, normalized to their respective linear averages. The classical channels are allocated at the upper frequency slots above the quantum channel. The plots assume a fixed attenuation of $\alpha=\SI[per-mode=symbol]{0.2}{\decibel\per\kilo\metre}$, a channel spacing of $\fch=\SI{50}{\giga\hertz}$, and a group-velocity-dispersion parameter of $\beta_{2}=\SI[per-mode=symbol]{-28}{\pico\second\squared\per\kilo\metre}$. As the channel count increases, with as few as 8 channels, the \gls*{fwm} evolution is closely represented by the approximation.

\subsection{Dependence on fiber length}

In the counter-propagating scenario, the effects accumulate up to a saturation level, whereas in the co-propagating case all contributions are attenuated by fiber loss for sufficiently long lengths.

Both \gls*{fwm} and co-propagating linear leakage generally decay with length, whereas the co-propagating \gls*{sprs} contribution initially grows and peaks at distance $\Lpk$. The peak length $\Lpk$ is obtained from the co-propagating \gls*{sprs} term in Eq.~\eqref{eq:cf} for the interference contribution from $f_{h}$ to $f_{i}$ as
\begin{equation}
  \left. \frac{\der}{\der L}\eta_{ih}P_{h}\frac{e^{(\alpha_{i}-\alpha_{h})L}-1}{\alpha_{i}-\alpha_{h}}e^{-\alpha_{i}L}\right|_{L=\Lpk}=0,
\end{equation}
after which, through algebraic manipulation, the expression for $\Lpk$ is obtained as
\begin{equation}
  \Lpk=\frac{1}{\alpha_{h}-\alpha_{i}}\ln\left(\frac{\alpha_{h}}{\alpha_{i}}\right),~\lim_{\alpha_{h}\rightarrow\alpha_{i}}\Lpk=\frac{1}{\alpha_{i}}.
\end{equation}

The co-propagating interference contributions evolve at distinct rates along the fiber, and an effective length can be defined for each contribution. The effective length is computed by integrating the corresponding contribution over the fiber length and normalizing it to its peak value. For the effects considered here, for each frequency contribution, the effective length at the quantum channel is given by
\begin{equation}
  \begin{split}
    L^{\mathrm{SpRS}}_{\eff}&=\frac{\alpha_{i}(1-e^{-\alpha_{h}L})-\alpha_{h}(1-e^{-\alpha_{i}L})}{\alpha_{i}\alpha_{h}(e^{-\alpha_{h}\Lpk}-e^{-\alpha_{i}\Lpk})},\\
    L^{\mathrm{FWM}}_{\eff}&=\frac{1}{2}\left[\frac{e^{(\Delta\alpha_{ihkl}-\alpha_{i})L}-1}{\Delta\alpha_{ihkl}-\alpha_{i}}+\frac{1-e^{\alpha_{i}z}}{\alpha_{i}}\right],\\
    L^{\mathrm{Linear}}_{\eff}&=\frac{1-e^{-\alpha_{i}L}}{\alpha_{i}},
  \end{split}
\end{equation}
where, for the \gls*{fwm} term, the approximation in Eq.~\eqref{eq:fwmavg} is used.

The counter-propagating \gls*{sprs} noise and the Rayleigh-backscattered linear noise grow with fiber length at similar rates, scaling as ${(1-e^{-(\alpha_i+\alpha_h)L})}$ and ${(1-e^{-2\alpha_i L})}$, respectively.

\begin{figure}[!t]
    \centering
    \includegraphics{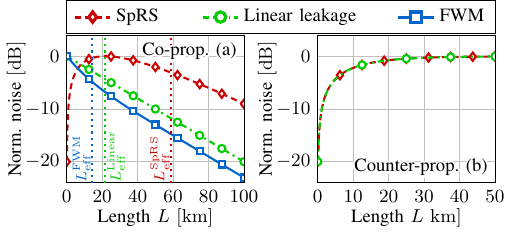}
    \caption{Normalized interference versus fiber length for the different sources (distinguished by markers) in (a) co-propagating and (b) counter-propagating scenarios.}
    \label{fig:Length}
\end{figure}

The normalized longitudinal dependence of all contributions depends only on attenuation. Figure~\ref{fig:Length} shows how each interference contribution evolves with fiber length, normalized to its peak value. Fiber attenuation is assumed constant and equal to ${\alpha=\SI[per-mode=symbol]{0.2}{\decibel\per\kilo\metre}}$ for this purpose. In Fig.~\ref{fig:Length}(a) we plot the evolution of the co-propagating interference. The \gls*{fwm} contribution is strongly concentrated at short distances, whereas \gls*{sprs} is dominant at long reachess, with the linear leakage contribution in between. Dotted lines indicate the asymptotic effective lengths (${L\rightarrow\infty}$) of the co-propagating contributions. Both counter-propagating contributions, shown in Fig.~\ref{fig:Length}(b), evolve at similar rates for the parameters considered, converging to a saturation level after $\sim$\SI{20}{\kilo\metre}.

\subsection{Wavelength dependence}

\begin{figure*}[!ht]
    \centering
    \includegraphics{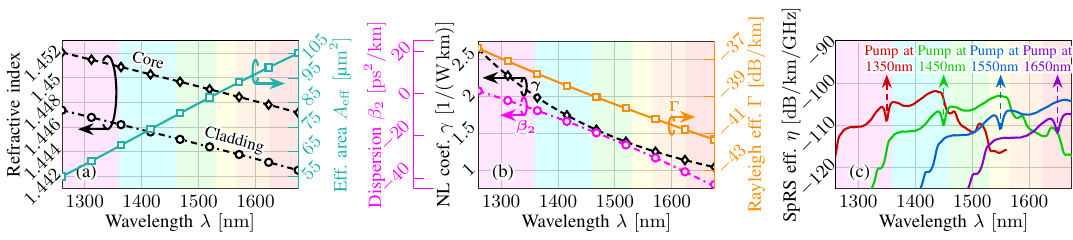}
    \caption{(a) Refractive index and effective area of the considered fiber. (b) Wavelength dependence of the group-velocity dispersion, non-linear coefficient, and Rayleigh efficiency. (c) \Gls*{sprs} efficiency for different pump wavelengths (color-coded). The shaded background regions indicate the transmission bands. From left to right, the colors correspond to the O-, E-, S-, C-, L-, and U-bands.}
    \label{fig:Design}
\end{figure*}

Many works have investigated multi-band transmission for coexistence schemes~\cite{townsend1997simultaneous,mao2018integrating,beppu2025coexistence,thomas2024quantum}, as non-linear interference can be significantly mitigated by increasing the spacing between classical and quantum channels. Nevertheless, fiber parameters can vary substantially between bands, requiring careful characterization.

Figure~\ref{fig:Design}(a) shows the frequency-dependent refractive index profile and effective area $A_{\mathrm{eff}}$ for a \SI{10}{\micro\metre}-core step-index fiber with a germanium-doped core and a core-to-cladding index ratio of \SI{0.31}{\percent}. The refractive index is calculated from Sellmeier's formula, with coefficients given in~\cite[Table~8]{murata1996handbook}.

In Fig.~\ref{fig:Design}(b), we plot the fiber's material group-velocity dispersion~($\beta_2$), non-linear coefficient~($\gamma$), and Rayleigh scattering efficiency~($\Gamma$). The non-linear coefficient and Rayleigh efficiency depend on the material properties as well as the refractive profile, and are considered to be $\SI{1.3}{\per\watt\per\kilo\metre}$ and $\SI[per-mode=symbol]{-40}{\decibel\per\kilo\metre}$ at $\SI{1550}{\nano\metre}$, respectively, scaling with wavelength as ${\gamma\propto(\lambda A_{\mathrm{eff}})^{-1}}$ and ${\Gamma\propto\lambda^{-4}}$~\cite{agrawal2007nonlinear}.

The dispersion is calculated from the derivatives of the mode propagation constant, typically crossing the axis within the O-band and increasing in magnitude with wavelength. As the dispersion coefficient increases, \gls*{fwm} is reduced, as is known and evident in Eqs.~\eqref{eq:fwmactual} and~\eqref{eq:dbetaidx}. Similarly, the non-linearity coefficient decreases as the wavelength and effective area increase, and the Rayleigh scattering efficiency decays rapidly with wavelength.

Figure~\ref{fig:Design}(c) shows the \gls*{sprs} efficiency for different pump channel wavelengths. With the exception of narrow peaks that matter in single-channel scenarios~\cite{thomas2024quantum}, \gls*{sprs} interference is overall a broadband effect. The curves plotted in Fig.~\ref{fig:Design}(c) are derived from the Raman gain efficiency provided in~\cite{lin2006raman,stolen1989raman}, using the analytical expression in~\cite[Eq.~(8)]{zischler2025accurate} to obtain the \gls*{sprs} efficiency, assuming the waveguide is at \SI{300}{\kelvin}. The \gls*{sprs} efficiency scales as ${\eta_{ih} \propto (\lambda_i A_{\mathrm{eff},i,h})^{-1}}$, where $A_{\mathrm{eff},i,h}$ is the effective cross-mode area between the lateral field profiles at $f_i$ and $f_h$~\cite[Eq.~(7)]{poletti2008description}, which varies non-negligibly across different bands.

The \gls*{sprs} efficiency decreases at anti-Stokes frequencies and as the pump wavelength increases. Therefore, allocating the stronger signals at longer wavelengths reduces interference in channels at shorter wavelengths, compared with the opposite configuration.

As observed in Fig.~\ref{fig:Design}, interference effects generally decrease as the wavelengths of the interference-inducing channels increase. Nevertheless, as discussed further, fiber attenuation also increases beyond its minimum in the C- and L-bands, indicating a non-trivial optimal frequency placement.

\subsection{Dependence on classical signals}

\begin{figure}[!b]
    \centering
    \includegraphics{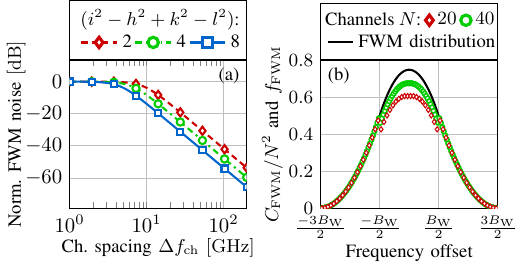}
    \caption{(a) Normalized \gls*{fwm} interference as a function of channel spacing for different index-mismatch values (${i^{2}-h^{2}+k^{2}-l^{2}}$). (b) Normalized \gls*{fwm} matching count $C_{\mathrm{FWM}}$ from~\eqref{eq:fwmcount} and the corresponding continuous distribution $f_{\mathrm{FWM}}$ from~\eqref{eq:fwmf}.}
    \label{fig:Bandwidth}
\end{figure}

The interference effects exhibit different dependencies on the classical signal parameters. For the \gls*{fwm} contribution, the propagation-constant mismatch $\Delta\beta_{ihkl}$ defines both the period of the longitudinal oscillations and the resulting interference intensity. For a fixed \gls*{wdm} grid, and using the approximation given in~\eqref{eq:dbetaidx} within Eq.~\eqref{eq:fwmavg}, the interference intensity is found to be proportional to both the channel-index mismatch and the channel spacing. Figure~\ref{fig:Bandwidth}(a) illustrates the reduction of \gls*{fwm} interference as a function of channel spacing, for different values of channel-index mismatch. The curves assume ${\alpha=\SI[per-mode=symbol]{0.2}{\decibel\per\kilo\metre}}$ and ${\beta_{2}=\SI[per-mode=symbol]{-28}{\pico\second\squared\per\kilo\metre}}$, and are normalized to their asymptotic peak values at ${\fch\rightarrow\SI{0}{\giga\hertz}}$. As the channel spacing increases, the \gls*{fwm} contribution decreases sharply, with a similar reduction observed for increasing channel-index mismatch. Consequently, increasing the separation between the quantum and classical channels by a few gigahertz can significantly suppress \gls*{fwm}. This result was presented in a similar form in~\cite[Fig.~1]{peters2009dense}.

As only channel combinations satisfying the phase-matching condition ${i-h+k-l=0}$ contribute to \gls*{fwm}, the overall interference intensity depends on the placement of the quantum channel in the \gls*{wdm} grid. Considering $N$ evenly spaced frequency channels within a total bandwidth $B_{\mathrm{W}}$, the number of \gls*{fwm} interference components affecting a channel with index $i$ is given by (derivation in~\appref{app:a})
\begin{equation}
  C_{\text{FWM}}=\begin{cases}
    \frac{N^{2}+N+2Ni+i+i^{2}}{2},&i<0,\\
    \frac{N^{2}-5N+2Ni-2i-2i^{2}+4 + 2|2i-N+1|}{2},&0\leq i<N,\\
    \frac{4N^{2}-2N-4Ni+i+i^{2}}{2},&N\leq i.
  \end{cases}
  \label{eq:fwmcount}
\end{equation}

As $N\rightarrow\infty$, Eq.~\eqref{eq:fwmcount} admits a continuous form, representing the combination count as a probability distribution normalized to the total bandwidth $B_{\mathrm{W}}$, given by
\begin{equation}
  f_{\mathrm{FWM}}=\begin{cases}
    \frac{1}{2}\left(\Delta_{i}+\frac{3}{2}\right)^{2},&\Delta_{i}<-\frac{1}{2},\\
    -\Delta_{i}^{2}+\frac{3}{4},&-\frac{1}{2}<\Delta_{i}<\frac{1}{2},\\
    \frac{1}{2}\left(\Delta_{i}-\frac{3}{2}\right)^{2},&\Delta_{i}>\frac{1}{2},\\
  \end{cases}
  \label{eq:fwmf}
\end{equation}
where $\Delta_i=\tfrac{f_i-f_c}{B_{\mathrm{W}}}$, with $f_c$ denoting the center frequency of the transmission band. The distribution is non-zero only within the interval $\Delta_i\in\bigl(-\tfrac{3}{2},\tfrac{3}{2}\bigr)$.

In Fig.~\ref{fig:Bandwidth}(b), we plot the normalized \gls*{fwm} density for different channel counts $N$, and the asymptotic probability distribution. For finite channel counts, the \gls*{fwm} intensity within the classical transmission band is reduced, since one classical signal must be removed to allocate the quantum channel. The distribution peaks at the center of the transmission band and decays by $\sim$\SI{33}{\percent} at the band edges, eventually decaying to zero at a separation of $B_{\mathrm{W}}$ from the classical signals.

The interference effects also exhibit distinct dependencies on channel power. While \gls*{sprs} and linear noise scale linearly, \gls*{fwm} scales with the third power of the channel power, as can be directly inferred from Eq.~\eqref{eq:cf}.

In the counter-propagating configuration, Rayleigh-backscattered linear noise and \gls*{sprs} grow at similar rates with channel power and length. The dominant effect is determined by the ratio of the \gls*{sprs} and Rayleigh efficiencies, as well as by the relative channel powers and the level of linear leakage. The linear leakage can be reduced by increasing the filter order and is often partially suppressed by the multiplexer, but it cannot be completely nullified.

\section{Quantum channel impairments}

Noisy photons impair the \glsxtrshort*{qkd} signal on different scales, depending on the protocol. In \gls*{dv-qkd}, information is encoded in the individual photons, resulting in a relatively low tolerance to noise. In contrast, \gls*{cv-qkd} protocols encode information in weak coherent states, which are transmitted at higher power levels than \gls*{dv-qkd} signals and therefore tolerates higher levels of interference. In this section, we evaluate the asymptotic \gls*{skr} limits under a noisy and lossy channel using BB84 and \gls*{gmcs} as representative examples of DV- and \gls*{cv-qkd} protocols, respectively.

We assume ideal \glsxtrshort*{qkd} transmitters and receivers, such that the resulting asymptotic limits are determined solely by the fiber-channel impairments. The noise is assumed to be randomly polarized, Gaussian distributed, and spectrally flat within the quantum channel bandwidth. Under these assumptions, the noise \gls*{psd} at frequency $f_i$ is given by
\begin{equation}
  \Sint{i}=\frac{\Pint{i}}{\Bch},
\end{equation}
where $\Bch$ denotes the bandwidth of the \gls*{wdm} grid slots.

In this work, the \gls*{skr} is defined as the secure key fraction (measured in bits per symbol), following conventional entropy-based definitions of theoretical \gls*{qkd} works~\cite{shor2000simple,weedbrook2012gaussian}. Assuming an ideal error-correction algorithm, the asymptotic rate of secret bits for a given scheme is then obtained by multiplying the \gls*{skr} by the symbol rate.

\subsection{In DV-QKD protocols}

In the original BB84 protocol, the transmitter sends a sequence of photons randomly alternating between two mutually unbiased bases, with information encoded in the photon state~\cite{bennet1984quantum}. At the receiver, photons are measured in randomly selected bases. Coincidence events from mismatched bases are discarded during post-processing.

Assuming an ideal, lossless, and noiseless \glsxtrshort*{qkd} transmitter and receiver, the resulting \gls*{qber} is given by (detailed in~\appref{app:b})
\begin{equation}
  \mathrm{QBER}=\frac{\Sint{i}\Rdet}{hf_{i}T_{i}+\Sint{i}\Rdet},
  \label{eq:qber}
\end{equation}
where $T_{i}=e^{-\alpha_{i}L}$ is the channel transmittance, and ${\Rdet=\Bdet\Tdet}$ is the collection factor that determines the fraction of received noise power integrated into a single detection bin\footnote{The factor $\Rdet$ can theoretically be zero, since amplitude-encoded signals (such as polarization-basis or time-bin BB84) can have negligible bandwidth, making it theoretically possible with perfect synchronization and no timing jitter to filter only signal photons. In practice, however, receivers integrate over a finite time window and thus unavoidably collect some coexistence-induced noise photons.}. Here, $\Bdet$ denotes the quantum receiver bandwidth, and $\Tdet$ is the single-photon detector measurement time window. The factor $\Rdet$ results in a linear scaling of the noise \gls*{psd} and, unless otherwise specified, we assume $\Rdet=1$ throughout this work for simplicity.

The asymptotic \gls*{skr} is related to the \gls*{qber} as~\cite{shor2000simple}
\begin{equation}
  \rdv=\frac{T_{i}}{2}\left[1-2H_{2}(\mathrm{QBER})\right],
\end{equation}
where $H_{2}(\cdot)$ is the binary Shannon entropy function, the factor $\tfrac{1}{2}$ accounts for the sifting step, and $T_{i}$ denotes the channel transmittance, accounting for all propagation losses.

\subsection{In CV-QKD protocols}

In \gls*{cv-qkd}, information is encoded in continuous quantum states. Specifically, in \gls*{gmcs}, the bit sequence is modulated in coherent Gaussian-distributed symbols~\cite{grosshans2003quantum}.

In calculations, power values are normalized to the shot-noise variance, which is proportional to the local oscillator power~\cite{laudenbach2018continuous}.

The excess noise comprises multiple contributions as detailed in~\cite{laudenbach2018continuous}, one of which is the coexistence-induced impairment. To isolate the effects of the coexisting classical signals, in this study, we disregard all other noise sources ($\xi_{i}=\xi_{\text{coex},i}$).

Under the assumptions of white, Gaussian distributed coexistence-induced noise and a flat receiver frequency response, the coexistence noise variance, expressed in shot-noise units, is given by (see~\appref{app:c})
\begin{equation}
  \xi_{\text{coex},i}=\frac{\Sint{i}}{2hf_{i}}.
  \label{eq:snu}
\end{equation}

In reverse reconciliation protocols, the asymptotic rate of secret bits per transmitted pulse is given by
\begin{equation}
  \rcv=\betaEC I_{AB}-\chi_{EB},
\end{equation}
where $\betaEC\leq 1$ is the error-correction reconciliation efficiency, $I_{AB}$ is the mutual information between the transmitter and receiver, and $\chi_{EB}$ is the Holevo bound between an ideal eavesdropper and the receiver.

\begin{figure}[!t]
    \centering
    \includegraphics{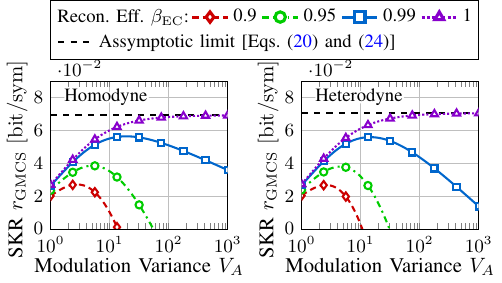}
    \caption{\Glsxtrlong*{skr} of \gls*{gmcs} for homodyne- and heterodyne-detected schemes as a function of modulation variance. The transmitter and receivers are assumed as ideal, lossless, and noiseless. Channel parameters are $T=\SI{-10}{\decibel}$ and $\xi=10^{-3}$.}
    \label{fig:Betas}
\end{figure}

Figure~\ref{fig:Betas} plots the \gls*{skr} versus modulation variance $V_{A}$ for both homodyne- and heterodyne-detected \gls*{gmcs}. With ideal reconciliation ($\betaEC=1$), the information rate and the Holevo bound increase at the same rate for large $V_{A}$, such that the \gls*{skr} approaches an upper bound as ${V_{A}\rightarrow\infty}$, as shown in the figure. Practical coding schemes cannot reach this asymptotic limit, but current coding schemes have reported efficiencies up to $0.99$~\cite{milicevic2018quasi,mani2021multiedge}. Assuming ideal transmitters and receivers, this asymptotic limit represents a theoretical upper bound for \glspl*{skr}, limited only by the lossy and noisy channel.

Assuming the signal as Nyquist-shaped with a symbol rate equal to the detector bandwidth (sinc-shaped pulses with zero roll-off), the asymptotic \gls*{skr} for $\betaEC=1$ and $V_{A}\rightarrow\infty$ in a homodyne-detected configuration is given by (derivation in~\appref{app:d})
\begin{equation}
  r_{\mathrm{GMCS}}=\frac{1}{2}\log_{2}\left[\frac{T_{i}\phi}{(\xi_{i}+1)(1-T_{i})}\right]-h_{2}(\omega),
  \label{eq:rhom}
\end{equation}
with $\phi$ given by
\begin{equation}
  \phi=T_{i}\omega+(1-T_{i})[2\omega-1+(\xi_{i}+\mu)/T_{i}],
\end{equation}
where $\mu=1$ for homodyne and $\mu=2$ for heterodyne detection, and where $\omega$ is given by
\begin{equation}
  \omega=\frac{\xi_{i}}{1-T_{i}}+1,
\end{equation}
and $h_{2}(\cdot)$ is given as
\begin{equation}
  h_{2}(x)=\frac{x+1}{2}\log_{2}\left(\frac{x+1}{2}\right)-\frac{x-1}{2}\log_{2}\left(\frac{x-1}{2}\right).
\end{equation}

For heterodyne detection, the asymptotic \gls*{skr} is given by
\begin{equation}
  \begin{split}
    r_{\mathrm{GMCS}}=&\log_{2}\left[\frac{T_{i}}{(\xi_{i}+2)(1-T)}\right]-h_{2}(\omega)\\
    &+h_{2}\left(\frac{\xi+2-T}{T}\right)+1-\log_{2}(e).
  \end{split}
  \label{eq:rhet}
\end{equation}

\section{Channel's SKR limits and interference mitigation}

\begin{figure}[!t]
    \centering
    \includegraphics{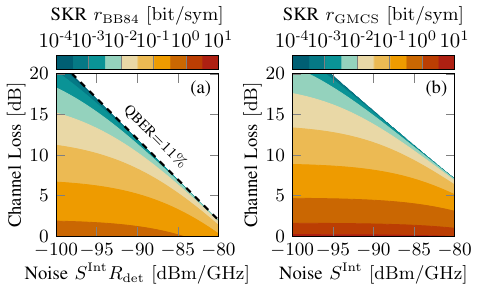}
    \caption{Asymptotic \glspl*{skr} as a function of channel loss and noise for (a) BB84 and (b) heterodyne-detected \gls*{gmcs} protocols.}
    \label{fig:SkrSweep}
\end{figure}

Under a lossy Gaussian-noise channel, the asymptotic limits of \gls*{gmcs} depend solely on the channel parameters, whereas the asymptotic limits of BB84 also depend on the collection factor. Figure~\ref{fig:SkrSweep} illustrates how the asymptotic \glspl*{skr} of both BB84 and \gls*{gmcs} vary with different levels of loss and noise. In Fig.~\ref{fig:SkrSweep}(a), it is highlighted the region where \gls*{qber} reaches the known threshold of $11\%$, above which no error-correction algorithm can recover the secret key~\cite{shor2000simple}. Both BB84 and \gls*{gmcs} scale similarly with respect to loss and noise, noting that \gls*{gmcs} achieve higher rates at low loss, as it can transmit more than a single bit per pulse.

\begin{figure}[!t]
    \centering
    \includegraphics{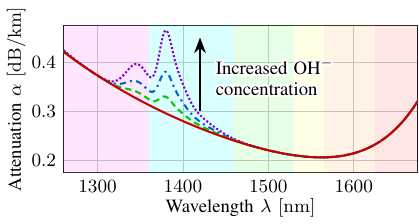}
    \caption{Fiber attenuation profiles for varying OH$^{-}$ concentrations, distinguished by line style. Curves are analytically calculated according to~\cite{walker1986rapid}.}
    \label{fig:Attenuation}
\end{figure}

In \gls*{wdm} classical-quantum links, the frequency dependence of both interference and attenuation affects \glsxtrshort*{qkd} protocols with different scaling. The frequency dependence of the interference effects was discussed in Section~\ref{sec:analytical}. While interference decreases at longer wavelengths, the attenuation profile of silica-core fibers exhibits a minimum in the C- and L-bands, as illustrated in Fig.~\ref{fig:Attenuation} for different water-peak levels.

\begin{figure}[!t]
    \centering
    \includegraphics{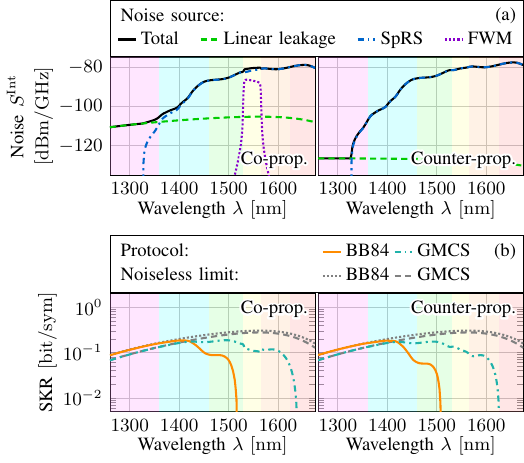}
    \caption{(a) Noise contributions from different sources and (b) asymptotic \glspl*{skr} considering the total interference in co- and counter-propagating scenarios, plotted alongside their respective loss-limited reference curves.}
    \label{fig:CBand}
\end{figure}

\begin{figure}[!t]
    \centering
    \includegraphics{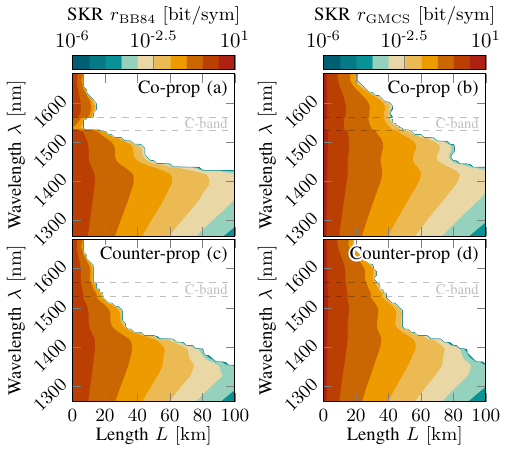}
    \caption{Asymptotic \glspl*{skr} as a function of the quantum channel wavelength and fiber length. \Glsxtrlong*{skr} curves are shown for (a,~c) BB84 and (b,~d) \gls*{gmcs} protocols, in both propagation directions.}
    \label{fig:SkrDistance}
\end{figure}

In Fig.~\ref{fig:CBand}(a), we plot the interference noise as a function of wavelength for a representative C-band-loaded scenario. Curves show the total noise and the individual contributions from each phenomena, calculated from Eq.~\eqref{eq:cf}. Noise within the transmission band assumes that the corresponding classical channel is deallocated. The curves consider a fiber length of ${L=\SI{25}{\kilo\meter}}$, and a channel spacing of ${\fch=\SI{50}{\giga\hertz}}$ with 88 polarization-multiplexed 16-QAM-shaped classical channels in the C-band. Each classical frequency channel is launched at ${P=\SI{-2}{\decibel m}}$, summed over both polarizations. Assuming a \SI{32}{\giga\baud} symbol rate with 0.1 roll-off factor, a flat-gain pre-amplifier with \SI{7}{\decibel} noise figure, and a minimum per-channel received power of \SI{-5}{\decibel m}, with~\cite[Eq.~(40)]{alvarado2018achievable}, the classical signals yield a combined \gls*{air} of \SI[per-mode=symbol]{22.53}{\tera\bit\per\second}. It is assumed a linear noise floor of ${\SI[per-mode=symbol]{-100}{\decibel m\per\giga\hertz}}$ at the classical transmitter.\footnote{This assumes a $\sim$\SI{30}{\decibel} filter at the quantum channel frequency, with unfiltered \gls*{ase} noise floor $\sim$\SI{20}{\decibel} below the transmission band \gls*{psd}, as measured from a reference C-band amplifier~\cite{zischler2026experimental}.} Frequency-dependent parameters are given in Fig.~\ref{fig:Design}, with attenuation corresponding to the OH$^{-}$-peak-absent case shown in Fig.~\ref{fig:Attenuation}.

The overall dominant noise source in both co- and counter-propagating scenarios is \gls*{sprs}. As expected from Fig.~\ref{fig:Design}(c), \gls*{sprs} noise is more significant in the L- and U-bands compared to the lower bands. Linear leakage is substantially mitigated by the Rayleigh factor in a counter-propagating configuration, whereas \gls*{sprs} remains at equal intensity. In the co-propagating configuration, \gls*{fwm} exhibits a non-negligible magnitude within the transmission band, comparable to the scale of \gls*{sprs}. Consistent with Fig.~\ref{fig:Bandwidth}(b), \gls*{fwm} is concentrated within the transmission band and negligible in the adjacent bands.

Figure~\ref{fig:CBand}(b) shows the asymptotic \glspl*{skr} of the BB84 and \gls*{gmcs} protocols derived from the total noise curves in Fig.~\ref{fig:CBand}(a). Within the O- and most of the E-band, the \glspl*{skr} reach the noiseless limit, indicating that noise impairments are negligible in this region. In this loss-limited region, BB84 and \gls*{gmcs} rates are similar. In the S- and upper bands, \gls*{gmcs} rates remain high and dropping near the U-band, as the protocol can tolerate increased noise, whereas BB84 rates drop sharply in the lower S-band.

Figure~\ref{fig:SkrDistance} shows the \glspl*{skr} as a function of wavelength and fiber length, for the same classical transmission configuration as in Fig.~\ref{fig:CBand}. Across protocols and propagation directions, the region below \SI{1400}{\nano\meter} remains unchanged, as coexistence noise is minimal and \glspl*{skr} are limited by the channel loss. For the given classical configuration, long distance \gls*{qkd} transmission is unachievable at higher wavelength.

\begin{figure}[!t]
    \centering
    \includegraphics{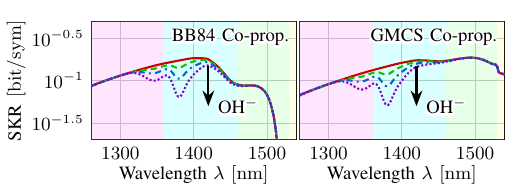}
    \caption{Asymptotic \glspl*{skr} for different OH$^{-}$ concentrations, with line styles corresponding to the attenuation curves shown in Fig.~\ref{fig:Attenuation}.}
    \label{fig:CBandAtt}
\end{figure}

Regardless of the protocol, Fig.~\ref{fig:CBand}(b) and Fig.~\ref{fig:SkrDistance} shows that \glspl*{skr} peak near \SI{1400}{\nano\meter}, within the E-band, for a given fiber length, due to the combination of lower noise and lower loss. As the water-peaks are located in the lower E-band, Fig.~\ref{fig:CBandAtt} illustrates how \glspl*{skr} vary in this region as the OH$^{-}$ concentration increases, from the attenuation curves plotted in Fig.~\ref{fig:Attenuation} for $L=\SI{25}{\kilo\meter}$.

The region near the OH$^{-}$ peaks may be undesirable for wide-scale deployment of \glsxtrshort*{qkd} services, as performance is highly dependent on fiber grade. Nevertheless, the upper E-band and lower S-band still provide higher \glspl*{skr} compared to the typical O-band, and are mostly unaffected by OH$^{-}$ concentration. These findings can be leveraged to improve \glspl*{skr} in future commercial coexistence deployments by tailoring \gls*{qkd} devices and components for operation within the upper E-band and lower S-band.

\begin{figure}[!b]
    \centering
    \includegraphics{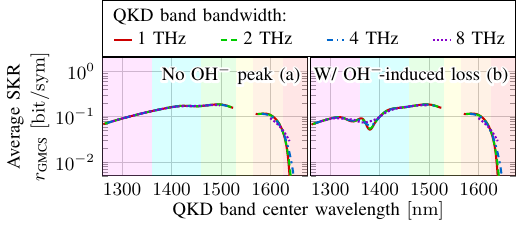}
    \caption{Asymptotic co-propagating \gls*{gmcs} \glspl*{skr} averaged over a \gls*{qkd} \gls*{wdm} band as a function of the band center wavelength. The curves in (a) consider the attenuation profile without the OH$^{-}$ absorption peak shown in Fig.~\ref{fig:Attenuation}, while (b) considers the corresponding profile with the highest absorption peak.}
    \label{fig:QkdBand}
\end{figure}

In Fig.~\ref{fig:QkdBand}, we plot the band-averaged asymptotic \gls*{gmcs} \gls*{skr} over a dedicated \gls*{qkd} band as a function of its center wavelength, for the considered bandwidths. The classical transmission parameters are as previously defined, and it is considered no overlap between the classical and \gls*{qkd} bands. Figure~\ref{fig:QkdBand}(a), which considers the attenuation profile without the OH$^{-}$ absorption peak, we observe overlap between the curves. In contrast, Fig.~\ref{fig:QkdBand}(b) shows that the sharp water-peak-induced drops are progressively smoothed as the \gls*{qkd} bandwidth increases. Overall, the behaviour of the \gls*{qkd} band follows that of a single channel. However, the increased bandwidth averages out localized drops, yielding a more uniform aggregate \gls*{skr}.

\begin{figure}[!t]
    \centering
    \includegraphics{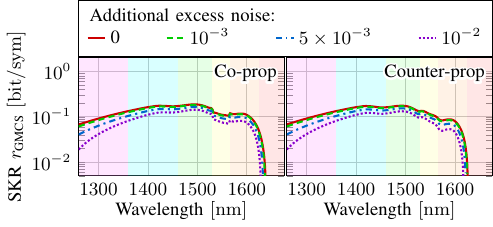}
    \caption{Asymptotic \gls*{gmcs} \glspl*{skr} in both propagating directions, accounting for the additional flat excess-noise values given in the legend.}
    \label{fig:AddNoise}
\end{figure}

As excess noise in \gls*{cv-qkd} arises from multiple sources, Fig.~\ref{fig:AddNoise} shows \glspl*{skr} under additional flat excess noise, simulating further impairments. This additional noise shifts the \glspl*{skr} for the given configurations, though they remain close to the limit set by coexistence-induced impairment alone.

\begin{figure}[!b]
    \centering
    \includegraphics{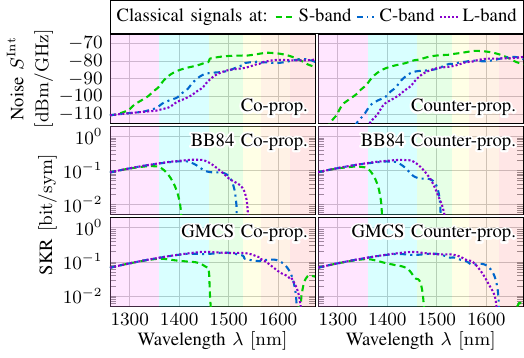}
    \caption{Co- and counter-propagating noise and asymptotic \glspl*{skr} for classical signals allocated in different transmission bands.}
    \label{fig:SCLBand}
\end{figure}

Finally, we consider scenarios in which classical traffic is transmitted in other bands than the C-band. Figure~\ref{fig:SCLBand} plots the noise spectral contributions for classical signals allocated in the S-, C-, or L-band and the corresponding \glspl*{skr}. The system parameters are the same as those for Fig.~\ref{fig:CBand}, with the L-band supporting 142 channels (\gls*{air}: \SI[per-mode=symbol]{36.35}{\tera\bit\per\second}) and the S-band 188 channels (\gls*{air}: \SI[per-mode=symbol]{48.13}{\tera\bit\per\second}). Interference increases from the C-band to the S-band due to higher total power and non-linearities. Notably, \gls*{sprs} from the S-band dominates across the entire E-band, significantly reducing \glspl*{skr}. Despite the increased channel count and power, interference from the L-band remains comparable to that generated by the C-band, due to reduced non-linearities. These findings indicate that expansion or reallocation toward the L-band notably improves the combined classical-quantum throughput, whereas S-band signals impose penalties even on O-band quantum channels.

\section{Conclusion}

Weak \glsxtrshort*{qkd} signals in classical-quantum \gls*{wdm} links are primarily affected by unfiltered linear leakage, \gls*{sprs}, and \gls*{fwm}. Each of these phenomena exhibits distinct spectral and longitudinal scaling and may become significant under different scenarios. Furthermore, the frequency dependence of fiber attenuation leads to non-trivial optimal configurations for \glsxtrshort*{qkd} channel allocation.

We analyze how the two-state BB84 and \gls*{gmcs} protocols scale asymptotically with noise. The presented analytical models enable quick assessment of the channel and identification of optimal configurations.

As demonstrated in representative scenarios and supported by the theory, improved \glspl*{skr} can be achieved in the upper E-band, where interference from the C-band is sufficiently suppressed and attenuation is lower than in the typically utilized O-band. The assertion sustains even when a continuous band is dedicated for \gls*{wdm} quantum channels. Additionally, allocating classical traffic to the L-band further suppresses \gls*{sprs} in the E- and lower S-bands. Contrary to conventional O-C quantum-classical coexistence schemes, E-L or S-L allocation can increase \glspl*{skr} without detriment to classical transmission rates.

\appendices
\def\sectionautorefname{appendix}

\section{Distribution of FWM contributions}
\label{app:a}

Considering a transmission band with $N$ evenly spaced channels. The interference channel indices are given by ${h,k,l\in[0,N-1]}$, while \gls*{fwm} interference is limited to channels ${i\in[-N+1,2N-2]}$. From the phase-matching condition (${i-h+k-l=0}$), the number of \gls*{fwm} contributions at a given index $i$, denoted $C_i$, where $k$ is determined by the remaining indices, is given by the count of pairs $(h,l)$ that satisfy
\begin{equation}
  i\leq h+l<i+N.
  \label{eq:a1}
\end{equation}

The number of combinations of $(h,l)$ that satisfy a given sum $x=h+l$, for $x\in(0, 2N-2)$, is
\begin{equation}
  C_{x}=\min(x+1,2N-x-1),
  \label{eq:a2}
\end{equation}
therefore, the total number of combinations $C_{i}$ that satisfy~\eqref{eq:a1} is given by
\begin{equation}
  C_{i}=\begin{cases}
    \sum_{x=0}^{N+i-1}x+1,&i<0,\\
    \sum_{x=i}^{N+i-1}\min(x+1,2N-x-1),&0\leq i<N,\\
    \sum_{x=i}^{2*N-2}2*N-x-1,&N\leq i,
  \end{cases}
  \label{eq:a3}
\end{equation}
where after simple manipulations, it can be expressed as
\begin{equation}
  C_{i}=\begin{cases}
    \frac{N^{2}+N+2Ni+i+i^{2}}{2},&i<0,\\
    \frac{N^{2}+N+2Ni-2i-2i^{2}}{2},&0\leq i<N,\\
    \frac{4N^{2}-2N-4Ni+i+i^{2}}{2},&N\leq i.
  \end{cases}
  \label{eq:a4}
\end{equation}

As the quantum signal is weak and therefore does not participate in the \gls*{fwm} interaction, contributions where ${h=i}$, ${k=i}$, and/or ${l=i}$ should be excluded. This only applies for ${0\leq i<N}$. The number of valid combinations that satisfy this criteria is
\begin{equation}
  C_{h=i}=C_{l=i}=N-1,~C_{k=i}=2\min(i, N-i-1).
  \label{eq:a5}
\end{equation}

After some manipulation, the number of \gls*{fwm} contributions in a dedicated quantum channel, excluding overlapping terms, is given by~\eqref{eq:fwmcount}.

\section{Asymptotic two-state BB84 SKRs}
\label{app:b}

Considering only photons at the correct encoding basis, the received photon rate is given by
\begin{equation}
  \Nrec=\etaDV T_{i}\Nsig,
\end{equation}
where $\etaDV$ is the receiver efficiency and $\Nsig$ is the transmitted \glsxtrshort*{qkd} photon rate.

The single-photon receiver uses a detection window of width $\tau_\mathrm{det}$, which, in the absence of timing jitter, should fully capture a single signal pulse. In time bin protocols, the measurement bases correspond to distinct time offsets within this window.

Interference photons measured within the detection window $\tau_\mathrm{det}$ are Poisson-distributed. The mean number of interference photons within the window $\tau_\mathrm{det}$ at frequency $f_i$ is proportional to the received noise power as
\begin{equation}
  \bar{N}^{\mathrm{Int}}_{i}=\frac{\Pint{i}\Bdet\tau_{\mathrm{det}}}{\Bch hf_{i}},
\end{equation}
where $\Bdet$ is the receiver's measurable bandwidth. Considering that one photon is transmitted per measurement interval (multiple-photon events should ideally be discarded~\cite{branciard2005security}), and neglecting optical shot noise, the mean number of received photon signal is $\etaDV T_{i}$, and the probability of a corrupted measurement is given by the \gls*{qber} as
\begin{equation}
  \mathrm{QBER}=\frac{\etaDV\bNint_{i}+\bNdc}{\etaDV T_{i}+\etaDV\bNint_{i}+\bNdc},
\end{equation}
where $\bNdc$ is the mean dark-count number, arising from thermal noise, electrical shot noise, and leakage at the quantum receiver. Optical shot noise can be neglected as it scales with the optical power and is therefore significantly weaker than the coexistence-induced interference and dark-count. Neglecting dark-count photons and assuming unitary receiver efficiency we obtain Eq.~\eqref{eq:qber}.

\section{Shot-noise units conversion}
\label{app:c}

The optical interference in \gls*{cv-qkd} protocols is quantified as the excess noise, normalized to the shot-noise variance. The shot-noise variance, expressed in volts squared, is given by~\cite[Eq.~(107)]{laudenbach2018continuous}  
\begin{equation}
  \sigma^{2}_{\mathrm{SN}}=\Plo\Bel\Rv^{2}hf_{i},
\end{equation}
where $\Plo$ is the local-oscillator power, $\Bel$ is the electronic bandwidth, and $\Rv$ is the receiver's conversion factor from optical power to volts, defined as the product of the photodetector responsivity and the transimpedance amplification gain.

Balanced photodetectors measure only photons that are mode-matched with the received local-oscillator photons, scaled by the receiver's conversion ratio. Therefore, in the respective polarization, the coexistence-induced excess noise in volts squared within the electronic bandwidth is given by
\begin{equation}
  \xi^{[\mathrm{V}^{2}]}_{\text{coex},i}=\frac{\Plo\Bel\Rv^{2}\Pint{i}}{2\Bch},
  \label{eq:snw}
\end{equation}
where we consider the noise \gls*{psd} as white, Gaussian, and constant within the receiver bandwidth, neglecting the receiver's frequency-dependent response. Normalizing Eq.~\eqref{eq:snw} to shot-noise units yields Eq.~\eqref{eq:snu}.

Derivations of the remaining excess noise sources appear in~\cite{laudenbach2018continuous}.

\section{Asymptotic GMCS SKRs}
\label{app:d}

The derivations provided in this section strongly reference the definitions given in~\cite[Section~III]{pirandola2024improved}.

The mutual information in \gls*{gmcs} protocols is given by 
\begin{equation}
  I_{AB}=\frac{\mu}{2}\log_{2}\left[1+\frac{\etaDV T_{i}V_{A}}{\etaDV\xi_{i}+\mu+\vel}\right],
\end{equation}
with $\mu=1$ for homodyne and $\mu=2$ for heterodyne detection.
where $\vel$ is the electronic noise.

The value of $\chi_{EB}$ is the Holevo bound between the eavesdropper and the receiver, and is given by
\begin{equation}
  \chi_{EB}=h_{2}(\lambda_{+})+h_{2}(\lambda_{-})-h_{2}(\tilde{\lambda}_{+})-h_{2}(\tilde{\lambda}_{-}),
  \label{eq:c1}
\end{equation}
where $\lambda_{\pm}$ are the symplectic eigenvalues of the eavesdropper state covariance matrix $\boldsymbol{\Sigma}_{E}$, and $\tilde{\lambda}_{\pm}$ are the same metric for the covariance matrix of the eavesdropper output state conditioned by the receiver's measurement $\boldsymbol{\Sigma}_{E|B}$.

\begin{figure}[!t]
    \centering
    \includegraphics{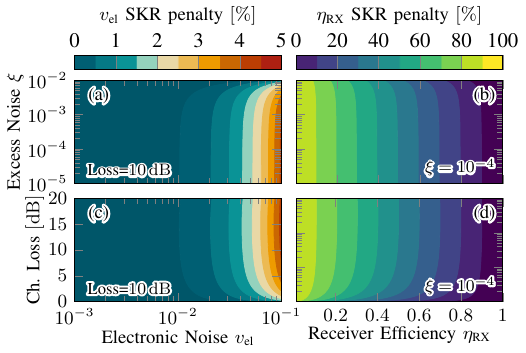}
    \caption{\Glsxtrlong*{skr} penalties arising from (a,c) non-zero electronic noise and (b,d) non-ideal receiver efficiency, obtained by sweeping the excess noise in (a,b) and channel loss in (c,d).}
    \label{fig:Penalty}
\end{figure}

The following subsections derive the asymptotic \gls*{skr} limits considering ideal reconciliation ($\betaEC=1$) and noiseless ($\vel=0$) and lossless ($\etaDV=1$) receiver. Penalties arising from non-ideal receiver devices are shown in Fig.~\ref{fig:Penalty} over a practical range of values. They increase nearly linearly with $\vel$ and decrease linearly with increasing $\etaDV$.

\subsection{Asymptotic limit derivations}

The symplectic eigenvalues can be obtained from the absolute ordinary eigenvalues of
\begin{equation}
  \lambda_{\pm}=\left|\eig\left(j\boldsymbol{\Omega\Sigma}_{E}\right)\right|,\quad
  \tilde{\lambda}_{\pm}=\left|\eig\left(j\boldsymbol{\Omega\Sigma}_{E|B}\right)\right|,
\end{equation}
where $\boldsymbol{\Omega}$ is a symplectic matrix, given for a 2-mode symbol by~\cite[Eq.~(2)]{weedbrook2012gaussian}
\begin{equation}
  \boldsymbol{\Omega}=\begin{bmatrix}
    0 & 1 & 0 & 0 \\
   -1 & 0 & 0 & 0 \\
    0 & 0 & 0 & 1 \\
    0 & 0 &-1 & 0
  \end{bmatrix}.
\end{equation}

The covariance matrix $\boldsymbol{\Sigma}_{E}$ is given by the block matrix
\begin{equation}
  \boldsymbol{\Sigma}_{E}=\begin{bmatrix}
    \omega\mathbf{I} & \psi_{1}\mathbf{Z}\\
    \psi_{1}\mathbf{Z} & \phi_{1}\mathbf{I}
  \end{bmatrix},
\end{equation}
where $\mathbf{I}$ is the identity matrix, ${\mathbf{Z}=\diag([1,-1])}$ with $\diag(\cdot)$ being the diagonal operator, and the remaining parameters are
\begin{equation}
  \begin{split}
    \psi_{1}&=\sqrt{T_{i}(\omega^{2}-1)},\\
    \phi_{1}&=T_{i}\omega+(1-T_{i})(V_{A}+1).
  \end{split}
\end{equation}

For a block matrix $\mathbf{M}$ of the following form
\begin{equation}
  \mathbf{M}=\begin{bmatrix}
    \diag([\omega_{A},\omega_{B}]) & \diag([\psi_{A},-\psi_{B}])\\
    \diag([\psi_{A},-\psi_{B}]) & \diag([\phi_{A},\phi_{B}])
  \end{bmatrix},
\end{equation}
the ordinary eigenvalues $\lambda_{M}$ of $j\Omega\mathbf{M}$ are obtained, from determinant properties,\footnote{Properties:\begin{itemize}
  \item $\left|\begin{bmatrix} \mathbf{A} & \mathbf{B} \\ \mathbf{C} & \mathbf{D} \end{bmatrix}\right|=|\mathbf{A}|\cdot|\mathbf{D}-\mathbf{C}\mathbf{A}^{-1}\mathbf{B}|$, if $\mathbf{A}$ is invertible.
    \item If $\mathbf{A}'$ equals $\mathbf{A}$ when a single row or column is swapped, ${|\mathbf{A}'|\Z{2}=\Z{2}-|\mathbf{A}|}$.
\end{itemize}} as the solution to
\begin{equation}
  \begin{split}
    \lambda_{M}=&\pm\sqrt{\frac{b'\pm\sqrt{b'-4c'}}{2}},\\
    b'=&\phi_{A}\phi_{B}+\omega_{A}\omega_{B}-2\psi_{A}\psi_{B},\\
    c'=&(\phi_{A}\phi_{B}-\psi_{A}\psi_{B})(\omega_{A}\omega_{B}-\psi_{A}\psi_{B})\\
    &-(\psi_{A}\phi_{B}-\omega_{A}\psi_{B})(\psi_{A}\omega_{B}-\phi_{A}\psi_{B}).
  \end{split}
  \label{eq:c2}
\end{equation}

For the covariance matrix $\boldsymbol{\Sigma}_{E}$, we have
\begin{equation}
  \begin{split}
    \omega_{A}&=\omega_{B}=\omega,\\
    \psi_{A}&=\psi_{B}=\psi_{1},\\
    \lim_{V_{A}\rightarrow\infty}\phi_{A}&=\lim_{V_{A}\rightarrow\infty}\phi_{B}=\lim_{V_{A}\rightarrow\infty}\phi_{1}\rightarrow(1-T_{i})V_{A},
  \end{split}
\end{equation}
and, after some manipulation, the symplectic eigenvalues are given by
\begin{equation}
  \lim_{V_{A}\rightarrow\infty}\lambda_{+}\rightarrow\phi,\quad\lim_{V_{A}\rightarrow\infty}\lambda_{-}=\omega.
\end{equation}

\subsubsection{Homodyne detection}

For homodyne detection, the parameters for the covariance matrix $\boldsymbol{\Sigma}_{E|B}$ are given by
\begin{equation}
  \begin{split}
    \lim_{V_{A}\rightarrow\infty}\omega_{A}&=\lim_{V_{A}\rightarrow\infty}\omega_{B}=\omega,\\
    \lim_{V_{A}\rightarrow\infty}\psi_{A}&=\psi,\quad\psi_{B}=\psi_{1},\\
    \lim_{V_{A}\rightarrow\infty}\phi_{A}&=\phi,\quad\lim_{V_{A}\rightarrow\infty}\phi_{B}\rightarrow(1-T_{i})V_{A},
  \end{split}
\end{equation}
with $\psi=\sqrt{(\omega^{2}-1)/T_{i}}$. The symplectic eigenvalues are then given by
\begin{equation}
  \lim_{V_{A}\rightarrow\infty}\tilde{\lambda}_{+}\rightarrow\sqrt{(1-T)V_{A}\phi},\quad\lim_{V_{A}\rightarrow\infty}\tilde{\lambda}_{-}=1.
\end{equation}

For the limit ${V_{A}\rightarrow\infty}$, the elements with $I_{AB}$, $h_{2}(\lambda_{+})$ and $h_{2}(\tilde{\lambda}_{+})$ in~\eqref{eq:c1} converge\footnote{$\lim\limits_{x\rightarrow\infty}h_{2}(x)\rightarrow\log_{2}(x)+\log_{2}(e)-1$} to the logarithm term in Eq.~\eqref{eq:rhom}.

\subsubsection{Heterodyne detection}

For heterodyne detection, the parameters for the covariance matrix $\boldsymbol{\Sigma}_{E|B}$ converge to their unsubscripted values (e.g. $\lim_{V_{A}\rightarrow\infty}\omega_{A}=\omega)$, and the symplectic eigenvalues are given by
\begin{equation}
  \lim_{V_{A}\rightarrow\infty}\tilde{\lambda}_{+}=\frac{\xi+2-T}{T},\quad\lim_{V_{A}\rightarrow\infty}\tilde{\lambda}_{-}=1.
\end{equation}

For the limit ${V_{A}\rightarrow\infty}$, the elements with $I_{AB}$ and $h_{2}(\lambda_{+})$ in~\eqref{eq:c1} converge to the logarithm term in Eq.~\eqref{eq:rhet}.

\bibliographystyle{IEEEtran}
\bibliography{references}

\end{document}